\documentclass[apjl]{emulateapj}
\newcommand{\Msun}{~M_\odot}

\newcommand{\kms}{\rm ~km~s^{-1}}

\begin{document}

\title{PULSAR WIND BUBBLE BLOWOUT FROM A SUPERNOVA}

    \author{John M. Blondin}
    \affil{Department of Physics, North Carolina State University,  Raleigh, NC 27695-8202}
    \email{blondin@ncsu.edu}
    \and 
    \author{Roger A. Chevalier}
    \affil{Department of Astronomy, University of Virginia, P.O. Box 400325, 
Charlottesville, VA 22904-4325}

\begin{abstract}

For pulsars born in supernovae, 
the expansion of the shocked pulsar wind nebula is initially
in the freely expanding ejecta of the supernova.
While the nebula is in the inner flat part of the ejecta density profile, the swept-up, accelerating shell is subject to the
Rayleigh-Taylor instability.
We carried out 2 and 3-dimensional simulations showing that the instability gives rise to filamentary structure during this initial phase
 but does not greatly change the dynamics of the  expanding shell.
 The flow is effectively self-similar.
If the shell is powered into the outer steep part of the density profile, the shell is subject to a robust Rayleigh-Taylor instability
in which the shell is fragmented and the shocked pulsar wind breaks out through the shell.
The flow is not self-similar in this phase.
For a wind nebula to reach this phase requires that the deposited pulsar energy be greater than the supernova energy,
or that the initial pulsar period be in the ms range for a typical $10^{51}$ erg supernova.
These conditions are satisfied by some magnetar models for Type I superluminous supernovae.
We also consider the Crab Nebula, which may be associated with a low energy supernova for which  this scenario applies.

\end{abstract}

\keywords{ISM: individual objects (Crab Nebula) --- stars: neutron ---supernovae: general}

\section{INTRODUCTION}

\begin{figure}[!hbtp]
\begin{center}
\includegraphics[width=4in]{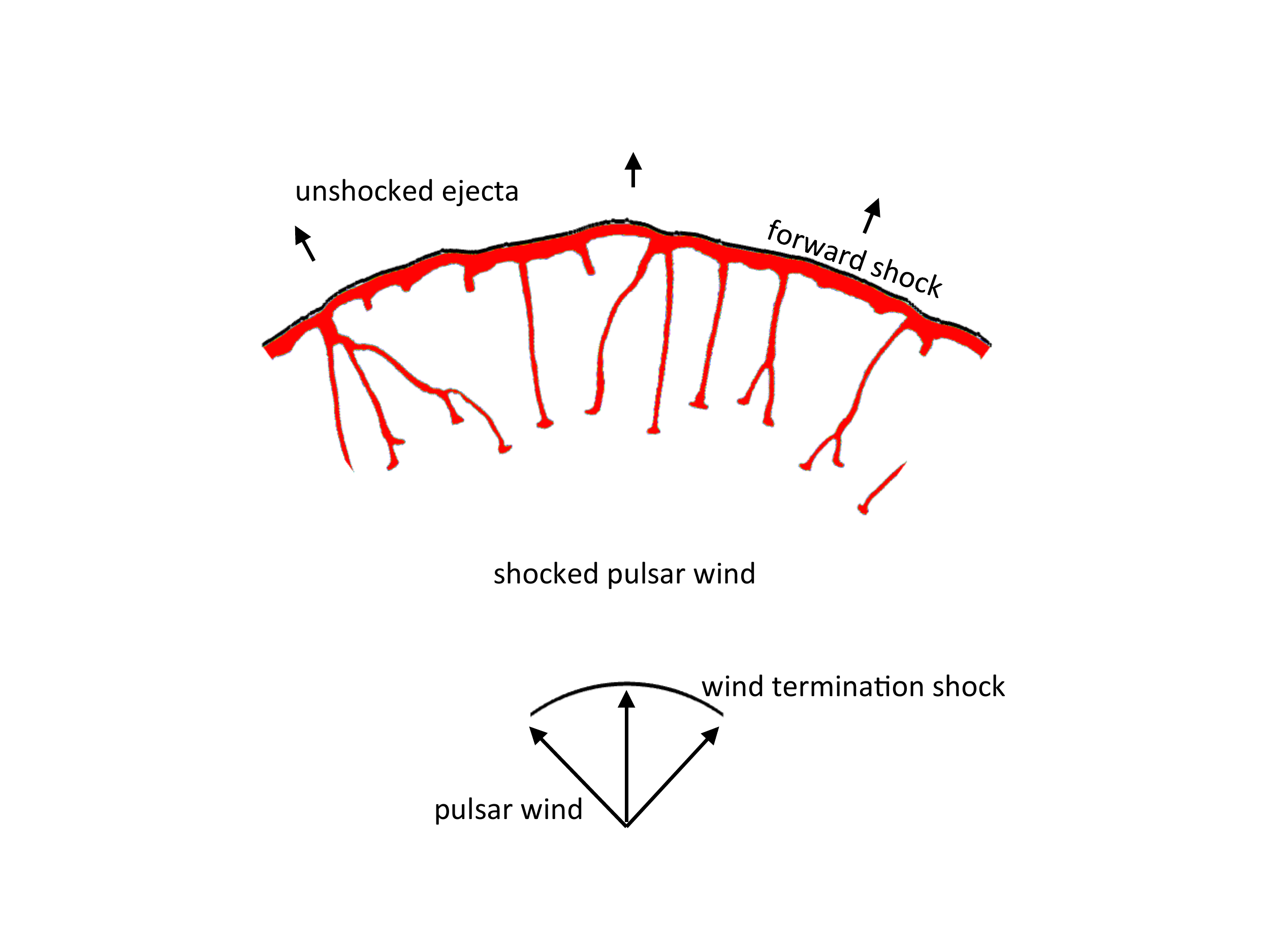}
\caption{Schematic diagram of a pulsar wind bubble within expanding supernova ejecta.
The supersonic pulsar wind is slowed and heated at the wind termination shock at a radius much smaller
than the forward shock.  A thin shell of shocked ejecta (shown in red) is swept up by the forward shock of
the wind bubble.  The acceleration of this shell leads to trailing fingers 
of dense ejecta gas in the interior of the wind bubble.}
\label{fig:schematic}
\end{center}
\end{figure}

A plausible model for the Crab Nebula involves the expansion of the wind bubble created by the Crab pulsar into the
freely expanding gas of the supernova \citep{chevalier77,hester08,bucc11}.
In this model, illustrated in Figure \ref{fig:schematic}, the expanding bubble of shocked pulsar wind 
sweeps up a thin shell of ejecta that accelerates in approximate accord with the
observed acceleration of Crab filaments \citep{trimble68}.
The shell is subject to the Rayleigh-Taylor instability (RTI), which provides an explanation for the filamentary
structure observed in the Crab Nebula.   
Numerical simulations have confirmed that the RTI can account for the filamentary structure
\citep{jun98,bucc04,porth14}.
The RTI gives rise to inner filaments that effectively broaden the swept up shell.

Supernova density profiles are expected to have an inner, flat region and an outer steep power law region;
the Crab Nebula is usually inferred to be interacting with the inner, flat region,
based on the low velocities that are observed in the Crab.
\cite{chevalier05} conjectured that the nature of the RTI would be qualitatively different depending on
whether the accelerating shell was in the inner flat or outer steep part of the density profile.
In the first phase, even if all the shell gas flows into filaments the ram pressure of the unshocked ejecta can contain the
pulsar wind nebula, albeit at a larger radius than in the case of a shell being present \citep{chevalier05}.
In the second phase,  the preshock ejecta gas is unable to confine the pulsar bubble by its ram pressure and 
there is the possibility of the pulsar nebula blowing out through the shell.
The approximate requirement for a nebula to make it to the second phase is that the energy deposited by the pulsar
be greater than the supernova energy, which  requires an initial pulsar rotation period in the ms range for a $10^{51}$ erg supernova.

\cite{bandiera83} suggested 
 that RTIs allow pulsar nebula emission to escape from a supernova, giving radio supernovae; however, radio emission is observed to be present from
 early times so the energetics are implausible and circumstellar interaction is a more likely source of the emission \citep{chevalier82}.
Although most pulsars are thought to be born with initial rotational energies $< 10^{51}$ ergs, there is the possibility of rapid initial rotation
for a small fraction of pulsars \citep{kotera13}.
This notion is exemplified by the millisecond magnetar model for superluminous supernovae (SLSNe) \citep{kasen10,woosley10}.
In this model, the pulsar power is assumed to be thermalized in the pulsar wind bubble and the eventual escape of the radiation gives
rise to a very luminous supernova.
Initial rotational energies of magnetars as high as $(1-2)\times 10^{53}$ ergs have been suggested \citep{metzger15}, so there
are possibilities for magnetar bubble blowout.
 \cite{kasen10} and \cite{kasen15} mentioned the likelihood of the RTI in the magnetar case; the consequences were described
as a broadening of the swept up shell, with little effect on the overall picture.
\cite{arons03} had earlier suggested that the bubble created by an energetic magnetar would break through the supernova ejecta
because of RTIs.
 \cite{chen16} recently carried  out 2-dimensional (2-D) simulations of magnetar bubbles that get into the blowout phase;
one of their models has an initial rotation rate of 1 ms and shows evidence for breakout through the swept-up shell.
\cite{suzuki16}  presented  a 2-D model  for the breakout of a bubble created by a relativistic wind.
Here, we investigate the qualitative difference in the RTI depending on whether the forward shock front
is in the inner, flat part of the supernova density profile or the outer, steep part of the profile.
The conditions for blowout are discussed in Section 2 and numerical simulations of the event, in 2-D and 3-D, are presented in Section 3.
A comparison to previous studies of the blowout phase is in Section 4.
The relevance to observed events is discussed in Section 5 and conclusions are in Section 6.

\section{CONDITIONS FOR BLOWOUT}
\label{sec2}

The initial supernova explosion with energy $E_s$ and ejecta mass $M_e$ is expected to reach free expansion on a timescale
of days or less, which is
less than  that for the pulsar bubble evolution.  The free expansion velocity distribution is $v=r/t$.
We consider a simple model for the supernova density distribution in which there is a flat inner power law and an outer steep power law \citep{chevalier89,matzner99},
as has also been assumed in calculations of magnetar bubble expansion \citep[e.g.,][]{kasen15}.
We have an inner profile $\rho_i =A t^{-3} (r/t)^{-\delta}$ and an outer profile  $\rho_o =B t^{-3} (r/t)^{-n}$, where $\delta <3$
(so  the mass does not diverge at small radius),
$n> 5$ (so the energy does not diverge at large radius), and $A$ and $B$ are constants.
For simplicity and to compare with previous work, we take $\delta = 0.$
The density is continuous across the transition point between the inner and outer profiles, leading to the velocity at the transition point
\begin{equation}
v_{tr}= \left[\frac{2(5-\delta)(n-5)}{(3-\delta)(n-3)}\frac{E_s}{M_e}\right]^{1/2}.
\end{equation}
For  $\delta=0$ and $n=7$, we have $v_t=4080\, E_{51}^{1/2}M_5^{-1/2}\kms$, 
where $E_{s51}$ is $E_s$ in units of $10^{51}$ ergs
and $M_5$ is $M_e$ in units of $5\Msun$.
If the pulsar provides a steady initial power $\dot E_0$, the radius of the swept up shell is \citep{chevalier05},   
\begin{equation}
R_s=\left[\frac{(5-\delta)^3(3-\delta)}{(11-2\delta)(9-2\delta)}\frac{\dot E_0}{4\pi A}\right]^{1/(5-\delta)}t^{(6-\delta)/(5-\delta)}
\label{radius}
\end{equation}
where the thin shell approximation has been made.
The pulsar nebula is assumed to be composed of an adiabatic, $\gamma=4/3$ gas.
The time for the shell to reach the transition velocity is  \citep{chevalier05}
\begin{equation}
t_{tr}=\frac{2(11-2\delta)(9-2\delta)(n-5)}{(3-\delta)(5-\delta)^2(n-\delta)}\frac{E_s}{\dot E_p}=f_1\frac{E_0}{\dot E_0},
\end{equation}
where values of $f_1$ for pairs of ($\alpha$, $n$) are: 0.75 (0, 7); 1.3 (1, 7), 1.5 (0, 12); and 2.5 (1, 12).
We take $f_1 =1.5$ as a representative value.
The time $t_{tr}$ is the end of the initial evolution in the flat density profile.
Observations of typical Galactic pulsar wind nebulae indicate that the shell is generally within the inner flat density profile  \citep{chevalier05},
implying a deposited energy less than the supernova energy.
This is supported by population studies of pulsars showing that most pulsars are born with relatively long periods, $\ga 100$ ms \citep{kaspi06}.
Provided $\delta <3$ the evolution of the shell radius $R\propto t^{\eta}$, with $\eta =(6-\delta )/(5-\delta)$.
In the limit $\delta \rightarrow 3$, we have $\eta =1.5$, which is the value expected for a shell with no external medium
\citep{ostriker71}.

These results are based on a 1-D thin shell model, in which all the swept up mass is in a thin, spherical shell.
The RTI can lead to  modification of the results, even when the shell is in the inner flat part of the density structure.
The instability can fragment the shell, leading to some of the gas lagging behind the outer shock in filaments.
\cite{chevalier05} found that if the shell is assumed to fragment, so that the pulsar bubble impinges directly on the
external freely expanding medium, the expansion has the same power law in time as the case with a shell present,
but with a higher velocity (by a factor of 1.25 for $\delta=0$). 
The assumption is extreme, but the result is illustrative.
In this case, the ram pressure due to the external medium is able to contain the pulsar bubble.
However, in the outer ejecta with $n>5$, the ram pressure of the external medium drops more rapidly than the pressure
in the expanding pulsar nebula, so that the pulsar can blow out through the swept up gas.
This situation would also apply to the interaction with a swept up shell and a low density exterior, as in the model of \cite{ostriker71}.
When this imbalance occurs, the blowout of the pulsar bubble through the shell can occur.

The initial rotational energy of the pulsar is $E_0 =(1/2)I\Omega^2 = 2\times 10^{52} P_{ms}^{-2}$ ergs, where $\Omega=2\pi /P$ is the spin rate, $P_{ms}$ is the pulsar period $P$
in units of ms and a neutron star moment of inertia, $I$, of $10^{45}$ g cm$^2$ has been assumed.
The spindown of the pulsar is not a completely solved problem, but the spindown power is generally taken to be of the form
\begin{equation}
\dot E_0=K \frac{B^2 R^6 \Omega^4}{c^3},
\end{equation}
where $K$ depends on the angle $\alpha$ between the magnetic and rotational axes, $B$ and $R$ are the pulsar magnetic
field and radius, and $c$ is the speed of light.
A force free model for the pulsar magnetosphere leads to $K=1+\sin^2\alpha$ \citep{spitkovsky06},
while a vacuum magnetic dipole model give $K=(2/3)\sin^2\alpha$.
A range of values of $K$ have been used in the literature on superluminous supernovae \citep{nicholl17}.  
The initial spindown timescale is
\begin{equation}
t_p=\frac {E_0}{\dot E_0} = \frac{Ic^3}{2KB^2 R^6\Omega^2} = 0.4 K^{-1}B_{14}^{-2}P_{ms}^2 {\rm~days},
\end{equation}
where $B_{14}$ is the pulsar magnetic field in units of $10^{14}$ G.
The sweep up model requires that $t_p > t_{tr}$ or the pulsar power would stop while the shell is in the region of shallow
density gradient and the flow would tend toward free expansion.
We have
\begin{equation}
\frac{t_{p}}{t_{tr}} = \frac{ E_0}{1.5E_s}=\frac{I\Omega^2}{3E_s}=13 E_{s51}^{-1} P_{ms}^{-2}.
\end{equation}
Evolving to the breakout phase depends on the supernova energy and the initial pulsar spin and, to some extent, the density structure of the supernova.
The requirement is
\begin{equation}
P_{ms}\la 4 E_{s51}^{-1/2}.
\end{equation}
The breakout phase can be achieved for plausible values of the parameters.

\section{SIMULATIONS}

Two-dimensional simulations have been carried out to show the growth of instabilities in pulsar wind nebulae 
\citep{jun98,bucc04,porth14,chen16,suzuki16}. \cite{jun98} and \cite{porth14} calculated global simulations for the Crab Nebula that showed filament formation and broadening of the shell.
The results showed some semblance  to the Crab. These simulations made the assumption that the pulsar bubble is expanding into freely expanding gas of uniform density, and did not evolve to the point where the shock wave moved into the region with a steep density profile. The limited effect of the instability is consistent with the analytic arguments given in Section 2.
\cite{chen16}  and \cite{suzuki16} simulated the early evolution of a pulsar wind nebula expansion and 
continued the evolution to interaction with a steep ejecta density profile.
We compare their results to ours in Section 4.

The aim here is to simulate the evolution of the RTI both in the early phase in the constant density region, and as the outer shock front moves into the outer steep part of the density profile. 
We use the code Virginia Hydrodynamics-1 (VH-1) that has previously been used to examine the late phases of evolution of a pulsar wind nebula, when it interacts with the externally generated reverse shock wave \citep{blondin01,temim15}. In particular, the simulation of \cite{temim15} shows the growth of RTI during the continued addition of pulsar power, but the reverse shock front returned to the pulsar nebula before the bubble expanded into the region of steep density drop. Those simulations used a supernova density profile with $\delta = 0, n = 9$; here we use $\delta = 0, n = 7$.  We use a radial grid with 768 uniformly spaced zones and an angular zone width of $\pi/2400$, giving roughly square zones in the vicinity of the swept-up shell.  
The angular extent of the grid covers $0.4\pi$ and is centered on the equator.  Periodic boundary conditions are used in the angular direction. 
The radial grid is expanded to track the shell to provide roughly constant resolution as the shell expands by several orders of magnitude.  

The supernova ejecta are treated as an adiabatic gas with a ratio of specific heats of $\gamma=5/3$.  The ejecta could be radiation-dominated in the early evolution of supernovae.  We ran identical simulations with $\gamma=4/3$ in the ejecta, which produced a slightly thinner, denser shell, but the overall behavior and morphology was not changed.
The pulsar wind gas has a ratio of specific heats of $\gamma=4/3$  so the shocked wind nebula acts as an adiabatic, relativistic fluid.
Fluid motions associated with the wind bubble are non-relativistic in the first phase of evolution.  

The pulsar power is injected at the inner boundary as a highly supersonic wind with a speed that produces a post-shock sound
speed in the wind bubble of $0.19E_{51}^{1/2}M_5^{-1/2} c$.  This is generally lower than the value of $c/\sqrt 3$ for a relativistic gas, 
which allows for a larger time-step and more efficient computation, yet is substantially higher than the ejecta velocity.  The pulsar power 
is assumed to be constant with time.

\begin{figure}[!hbtp]
\begin{center}
\includegraphics[width=3.5in]{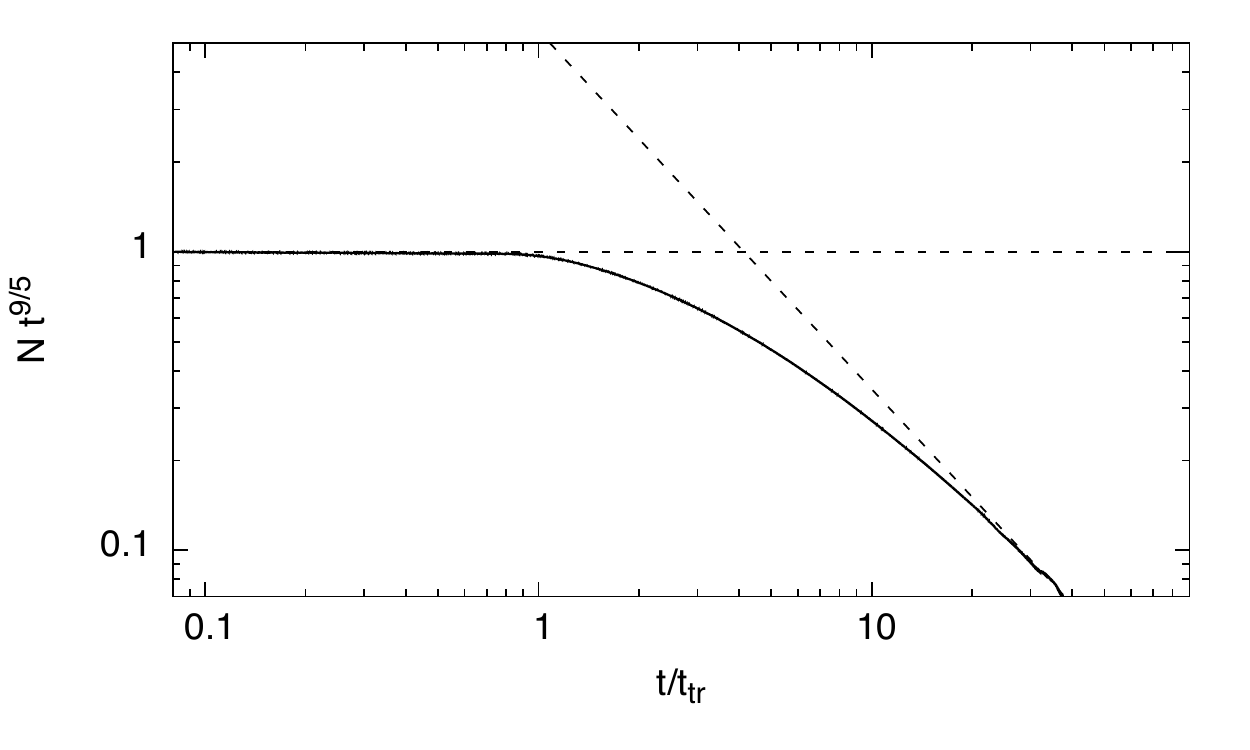}
\caption{Evolution of the radial column density in the spherically-symmetric (1-D) hydrodynamic simulation illustrating the 
sharp transition at $t\sim t_{tr}$. The horizontal dashed line corresponds to an evolution decreasing as $t^{-9/5}$ and the other
dashed line to an evolution decreasing as $t^{-3}$.  The vertical axis has been scaled to unity in the early self-similar phase.}
\label{fig:onedim}
\end{center}
\end{figure}

The evolution of a spherically-symmetric (1-D) numerical simulation is illustrated in Figure \ref{fig:onedim}.  
The simulation is started at an early time of $10^{-4}\, t_{tr}$ with a 
supersonic wind on the inner half of the grid and the expanding supernova ejecta on the other half.  
By the early times shown in Figure \ref{fig:onedim} the swept-up shell has already reached the initial self-similar state where
the radius of the shell follows the analytical result  in equation (\ref{radius}).  The radial column density in this
initial phase, $t < t_{tr}$,  is decreasing as $M/R^2 \propto t^{-9/5}$.  
Entering the late $t > t_{tr}$ phase, the shell is further accelerated and 
the expansion evolves toward another self-similar state, with shell radius $R\propto t^{1.5}$ \citep{ostriker71}.   
In this  phase 
the column density approaches an asymptotic power law decrease of $M/R^2 \propto t^{-3}$. 

\begin{figure*}[!hbtp]
\begin{center}
\includegraphics[width=6in]{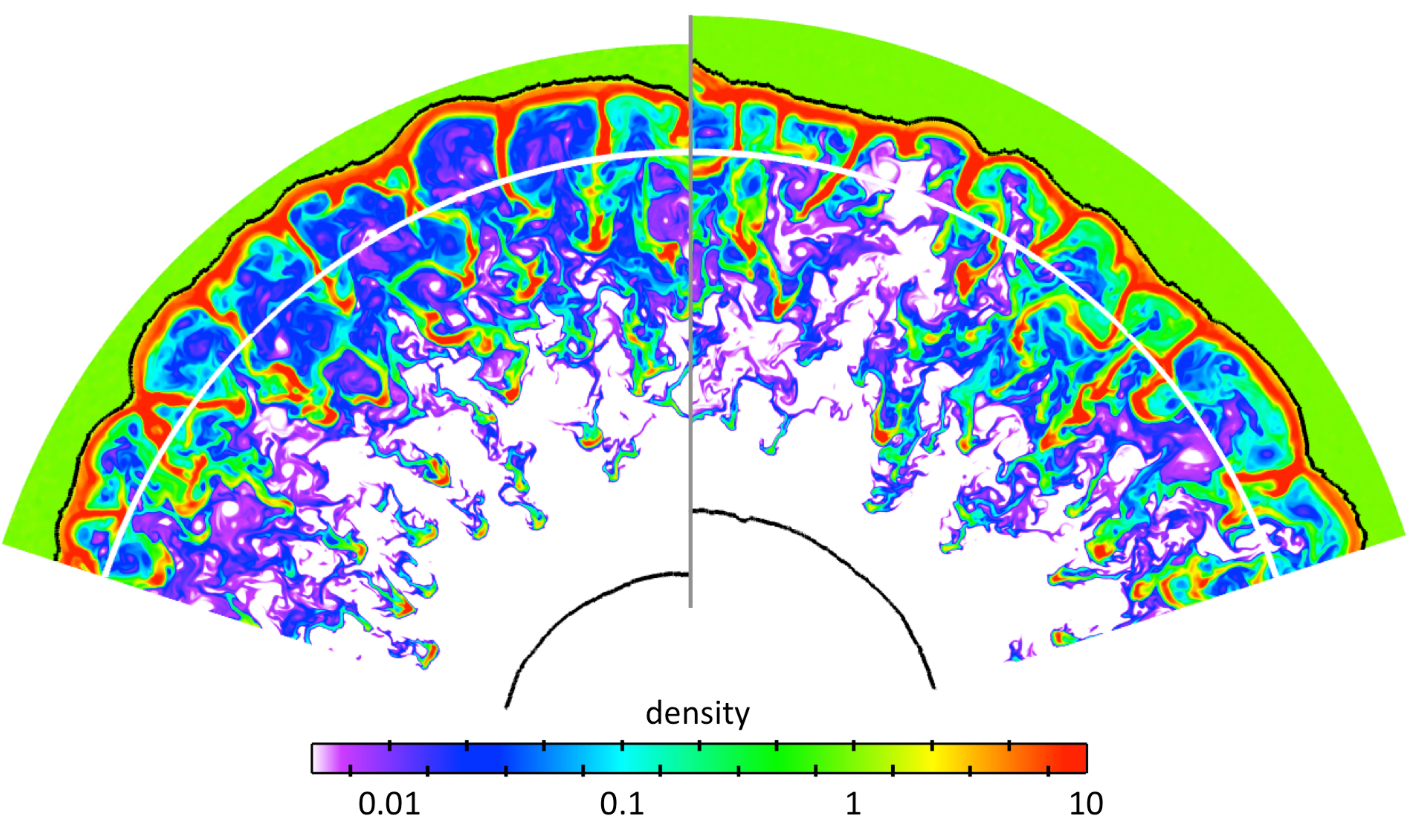}
\caption{Evolution of the RTI in the initial phase once the instability has reached a self-similar state.  The color
corresponds to the gas density normalized to the pre-shock density (in the plateau section of the ejecta) 
in a two-dimensional simulation on a $0.4\pi$ wedge.  The two frames correspond to times that differ by a factor of ten.
The images are scaled to put the location of the forward shock in a spherically symmetric model (marked by a thick white line) 
at the same radius.
The black lines outline the forward and wind termination shocks .}
\label{fig:self-sim}
\end{center}
\end{figure*}

The termination shock does not expand in a self-similar way with the main shell; it expands outward in the self-similar frame.
The position of the termination shock is determined by the distance it takes to slow the post termination shock flow to the
velocity of the shell.
In a shocked pulsar bubble with sound speed $\gg$ gas velocity, we have $p$ and $\rho$ about constant, while velocity $\propto r^{-2}$.
The pressure during the early phase evolves as $t^{-13/5}$ so the radius of the termination shock of the steady pulsar wind $R_{term}\propto t^{1.3}$,
versus $R_s \propto t^{1.2}$.
If $v_{term}$ is the velocity immediately downstream of the termination shock (fixed by the wind speed for a high Mach number shock) 
and $V_s$ is the shell velocity,
we have $R_s/R_{term}\approx (v_{term}/V_s)^{1/2}$; $V_s$ increases with time, so the shocked region becomes narrower.  For our
choice of pulsar wind speed we find $R_s/R_{term} \approx 3 (t/t_{tr})^{-0.1}$ for $t<t_{tr}$, where the coefficient is determined from a 1-D simulation.
This argument breaks down once the bubble gas can stream out through the ejecta.

The multidimensional simulations were initialized using the evolved 1-D profiles.  To examine the RTI in the first phase of evolution we initialized 
two-dimensional simulations at a time of $10^{-3}\, t_{tr}$ and continued the evolution to $20\, t_{tr}$ without the power-law region (i.e., continuing
the plateau region indefinitely).  Without any perturbations, the swept up shell began to show evidence for instability 
only after $10^{-1}\, t_{tr}$.  With 5\% perturbations in the preshock ejecta, the growth of Rayleigh-Taylor fingers begins earlier and reaches a 
larger amplitude by $t \sim t_{tr}$.
Perturbations in the ejecta can be expected because of instabilities   during the explosion phase and the growth of Nickel bubbles. 
In previous work, \cite{jun98} included 1\% perturbations in the ejecta to seed the instability. Porth et al. (2014) did not include any seed perturbations, but allowed for asymmetry in the power input by the central pulsar. The asymmetry drove turbulent motions in the bubble and the growth of the shell instability. 

While the forward shock front is in the flat part of the supernova density profile, the flow remains roughly self-similar.
Allowing 2-D motion does not introduce any new dimensional parameters and the breakup of the shocked shell by the RTI  does not
allow blowout through the shell; the preshock medium provides enough ram pressure to contain the pulsar bubble \citep{chevalier05}.
Figure \ref{fig:self-sim} shows the density structure at two times that are in the early self-similar phase but late enough that the RTI has saturated.
It can be seen that the structure remains qualitatively the same, although there are differences in detail.
A similar situation occurs for the deceleration of supernova ejecta by a low density surroundings \citep{chevalier92a}.
In the pulsar nebula, the swept up ejecta goes into the filaments and the shell; about 60\% of the total swept up mass is in the quasi-spherical shell at any
time in this self-similar phase, with the remainder in the dense filaments filling the bubble interior.
The outer shock wave radius has the expected $t^{6/5}$ time dependence and has a value $1.07 R_s$.
The expectation from the analytic theory is that the coefficient of $R_s$ would be between 1 and 1.25 \citep{chevalier05}.
The termination shock remains nearly spherical, but creeps forward with respect to the forward shock as discussed above.  

The extent of the RTI region is similar to that found by \cite{porth14}.
\cite{porth14} included a magnetic field in their simulations, but found that cases with high and low magnetization evolved similarly
with respect to the RTI.  These results suggest that a magnetic field is not a crucial factor in the development of the instability.
\cite{porth14} also carried out a study of the effect of resolution on the results, showing that going to high resolution primarily leads to
greater structure in the Kelvin-Helmholz instability that is related to the RTI.
A similar situation is present for the RTI driven by the deceleration of supernova ejecta \citep{chevalier92a}.
\cite{chen16}  carried out 2 simulations with resolution higher than ours.
The result is again more detailed structure resulting from the Kelvin-Helmholz instability \citep[see Fig.\ 5 of][]{chen16}.
We conclude that our simulations are adequately resolved for the larger scale structure.

In our simulations, the RTIs generate pressure waves that move in toward the pulsar wind termination shock.
The waves cause gentle deformation of the termination shock, but do not disrupt the shock front.
In the 2-D simulations of \cite{porth14}, \cite{chen16} and  \cite{suzuki16}, a filament forms along the symmetry axis that moves in to the termination shock.
By calculating just a wedge of the flow and avoiding any coordinate singularity, we have avoided this feature.
In their simulation of pulsar nebulae, \cite{camus09}, and \cite{porth14} found that there was feedback between the RTI in the outer part
of the nebula and the wind termination shock.
We  did not find such an effect, but did not allow for asymmetric power input from the pulsar wind as in \cite{camus09} and \cite{porth14}.
In any case, our results for the RT unstable region appear to be similar to those of the other authors.

\begin{figure}[!hbtp]
\begin{center}
\includegraphics[width=3.5in]{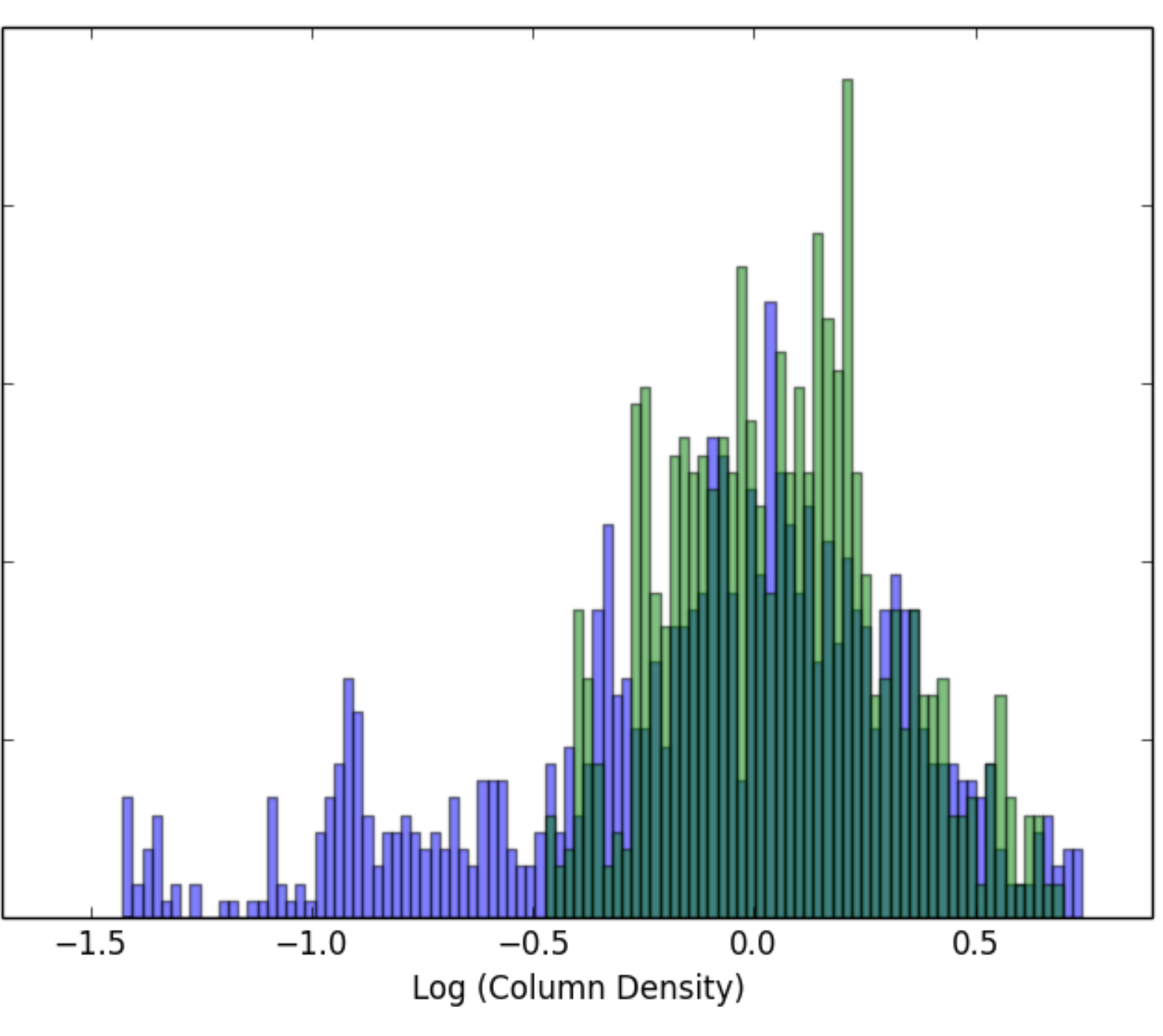}
\caption{Distribution of radial column densities normalized to the spherically symmetric solution.
The transparent green histogram represents the self-similar state of the RTI (corresponding to the left frame in Figure 2).
The blue histogram represents the blow-out phase at $t = 7 t_{tr}$ (corresponding to Figure 6).}
\label{fig:column_ss}
\end{center}
\end{figure}

There is some gentle deformation of the leading shock - which helps feed the long fingers - but the overall shell is still roughly spherical with 
relatively uniform column density.  Figure \ref{fig:column_ss} shows a histogram of the radial column
densities during the self-similar phase.  The shell itself is relatively uniform; less than 2\% of the shell has
a column density reduced by more than a factor of two.  
The fraction of lines-of-sight with significantly enhanced 
column density due to RTI fingers is less than a few percent, and no radial lines have substantially reduced column density relative to 
the spherical model.

\begin{figure}[!hbtp]
\begin{center}
\includegraphics[width=3.5in]{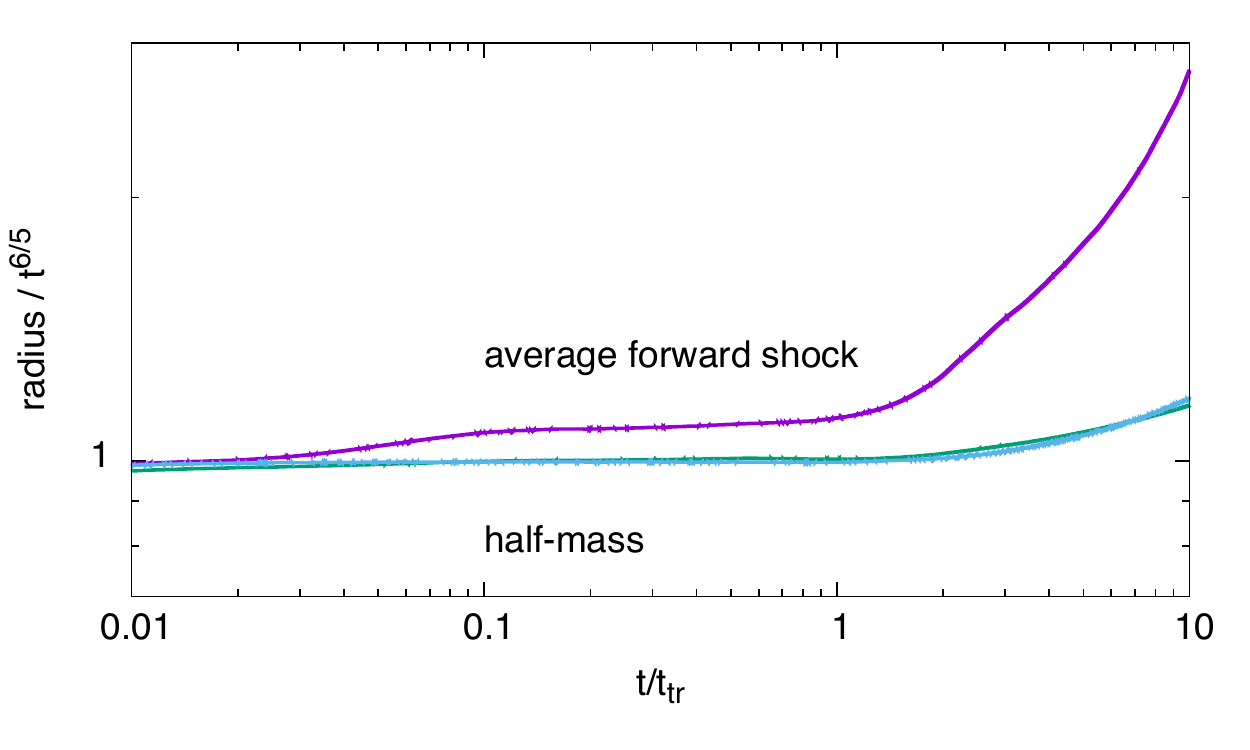}
\caption{Evolution of the forward shock relative to the radius enclosing half of the swept-up ejecta mass in a 2-D simulation.
At early times the swept-up shell is nearly spherical and the half-mass radius is very near the forward shock.  These two radii
diverge slightly due to the development of the RTI in the plateau phase.  Once this instability is saturated the evolution is again
self-similar.  Blowout occurs shortly after $t_{tr}$ and results in the forward shock accelerating relative to the bulk of the swept-up ejecta mass.}
\label{fig:notSS}
\end{center}
\end{figure}

We continued the pulsar power into the regime where the shell moved beyond the flat part of the supernova density profile. 
The instability proved to be robust in this phase. Even in the case of little perturbation growth in the initial phase, 
there was strong fragmentation of the shell once the shell entered the region of large density gradient.  
The simulations described here were initialized at $10^{-3}\, t_{tr}$, providing sufficient time for the growth of RTI fingers to reach saturation by 
the time the shell reaches the ejecta shelf.  Shortly after $t\sim t_{tr}$ the growth of the RT fingers is dramatically increased relative to 
the initial phase.  After a few $t_{tr}$ the shell is blown outward relative to the RT fingers, resulting in an evolution that is no longer
self-similar.  This deviation from an 'averaged self-similar' evolution is shown in Figure \ref{fig:notSS} where the angle-averaged forward shock radius is
compared to the radius enclosing half of the swept-up ejecta mass.  Once the instability in the early plateau phase is saturated ($\sim 0.1 t_{tr}$), the ratio
of these radii is roughly constant.  Once blowout occurs at $\sim t_{tr}$ the forward shock is rapidly accelerated relative to the bulk of the shocked
ejecta mass; the evolution in this later phase is not self-similar.

\begin{figure*}[!hbtp]
\begin{center}
\includegraphics[width=7in]{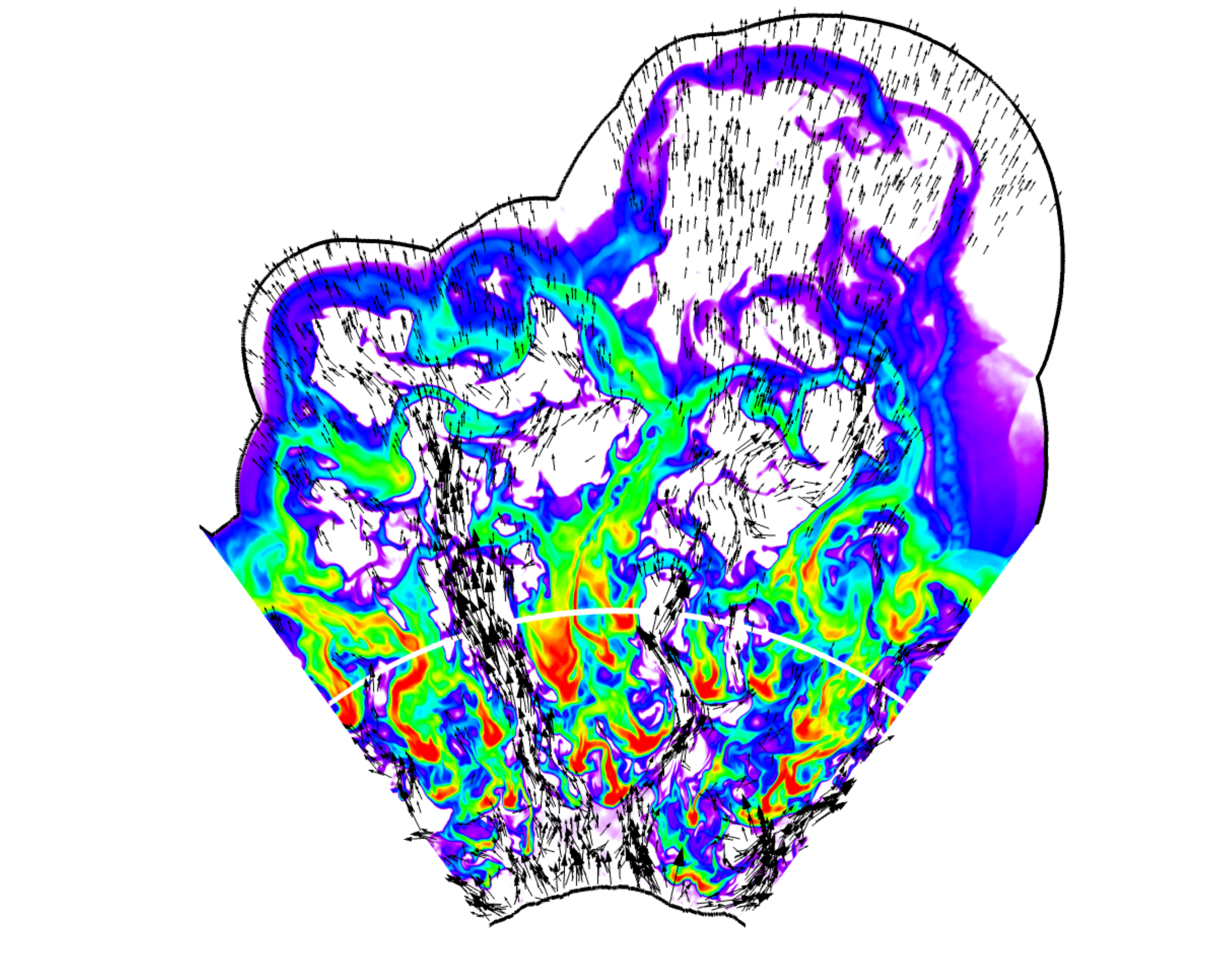}
\caption{The formation of channels in the blowout phase in a two-dimensional simulation on a $0.4\pi$ wedge at a time of 7 $t_{tr}$.
Only gas flow faster than the ejecta speed at the forward shock is shown for clarity and the high speed of the unshocked pulsar wind is not shown. 
The color depicts the logarithm of the gas density using the same color scheme and dynamic range of a factor of 2000 as in Figure 3, but with the scale adjusted
to the maximum density in the clumps.  The white line shows the forward shock radius in a 1-D spherically symmetric simulation with the same parameters. }
\label{fig:blowout}
\end{center}
\end{figure*}

In 2-D, the result is a very non-spherical forward shock marked by large bubbles from the blowout as shown in Figure \ref{fig:blowout}.
Roughly half the ejecta mass is left behind at radii less than the shell radius obtained in the spherically symmetric evolution.
The remnants of some RTI fingers can be seen close to the wind termination shock in, slightly perturbing
this shock from its initial spherical shape.  This termination shock interaction does not appear to affect the overall morphology of the blowout; 
similar simulations where the termination shock is forced to be spherically symmetric produce qualitatively similar results.
The regions of maximum blowout have negligible mass in the immediate post-shock shell.  Moreover, the blowout is associated
with evacuated channels of high-speed flow through the RTI fragments.  The resulting holes in the swept up ejecta are 
evident in the distribution of radial column densities shown in Figure \ref{fig:column_ss}.  The 
column densities in these blowout regions are up to an order of magnitude smaller than the spherical model at the same age.

\begin{figure*}[!hbtp]
\begin{center}
\includegraphics[width=6in]{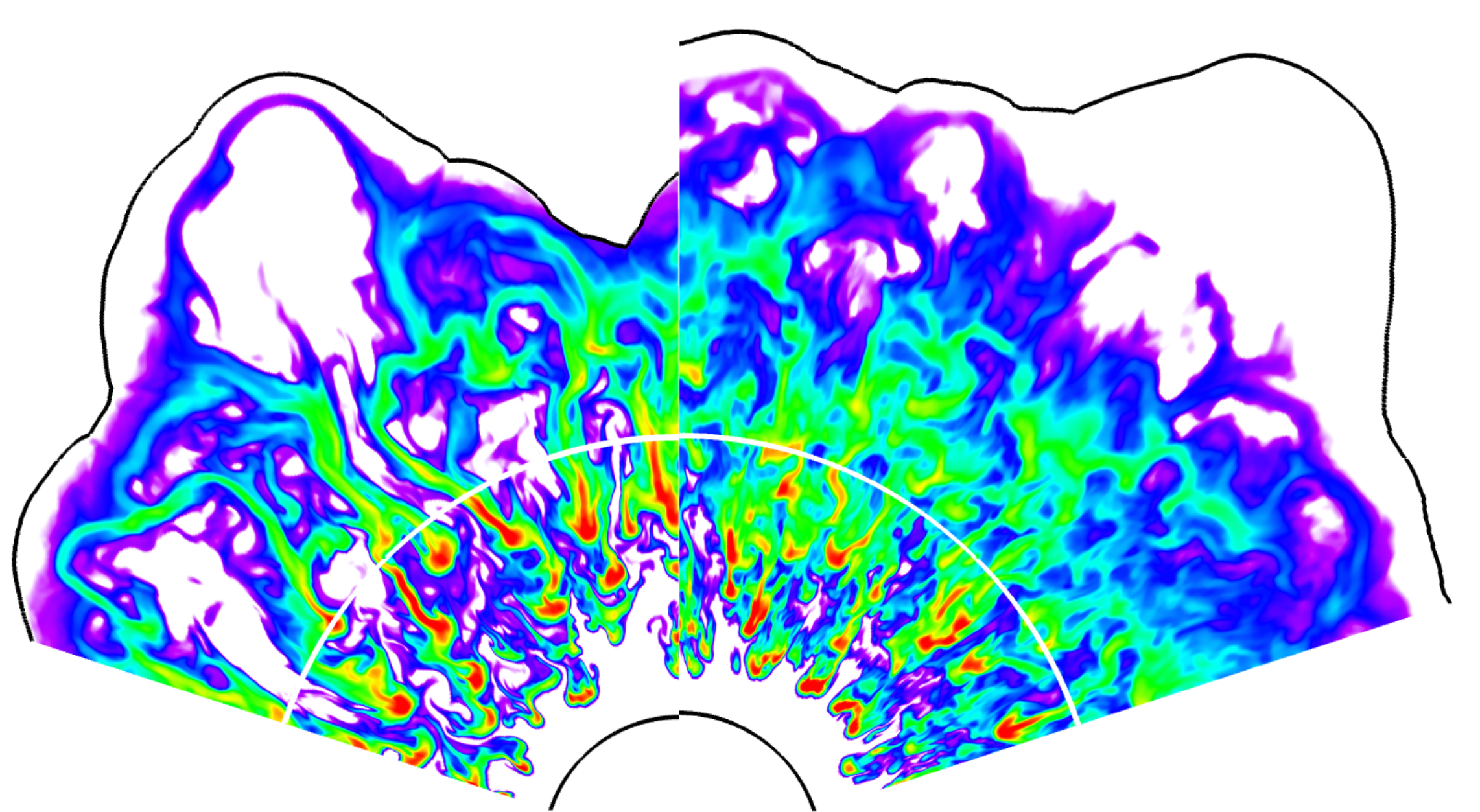}
\caption{The gas density in the blowout phase at a time of $7t_{tr}$ for a two-dimensional simulation (left) and
in a slice from a three-dimensional simulation (right). The color scale covers a factor of 2000 as in Figure 3, with the maximum
adjusted to the highest density in the clumps. The white line shows the forward shock radius in a 1-D spherically symmetric simulation.}
\label{fig:3dslice}
\end{center}
\end{figure*}

The behavior is qualitatively similar in three dimensions,  as shown in Figure \ref{fig:3dslice}.  The three-dimensional simulation and the 
corresponding two-dimensional simulation shown in Figure \ref{fig:3dslice} use half the spatial resolution of the previous models: 384 radial
zones and 480 angular zones covering $0.4\pi$.  At this lower resolution the RTI fragments do not reach the termination shock.  We therefore
chose to impose a spherical wind termination shock in order to improve computational speed and minimize numerical noise at this shock.
The instability is dominated by slightly higher order modes in three dimensions and the small-scale structure fills in the 
otherwise empty channels seen in two-dimensions.
Despite the less structured channels, the low density gas does largely escape through the dense shell and produces
comparable blow-out structures as seen in two dimensions.  One again finds half the ejecta mass inside the spherically-symmetric shell radius 
and a forward shock at roughly twice this radius.
The 1D evolution of an accelerated shell with $R_s\propto t^{1.5}$ is not maintained in the later evolution.
While the shell evolution is roughly self-similar in the first phase, this is not the case for the later phase.

\begin{figure}[!hbtp]
%\begin{figure}[here]
\begin{center}
\includegraphics[width=3.5in]{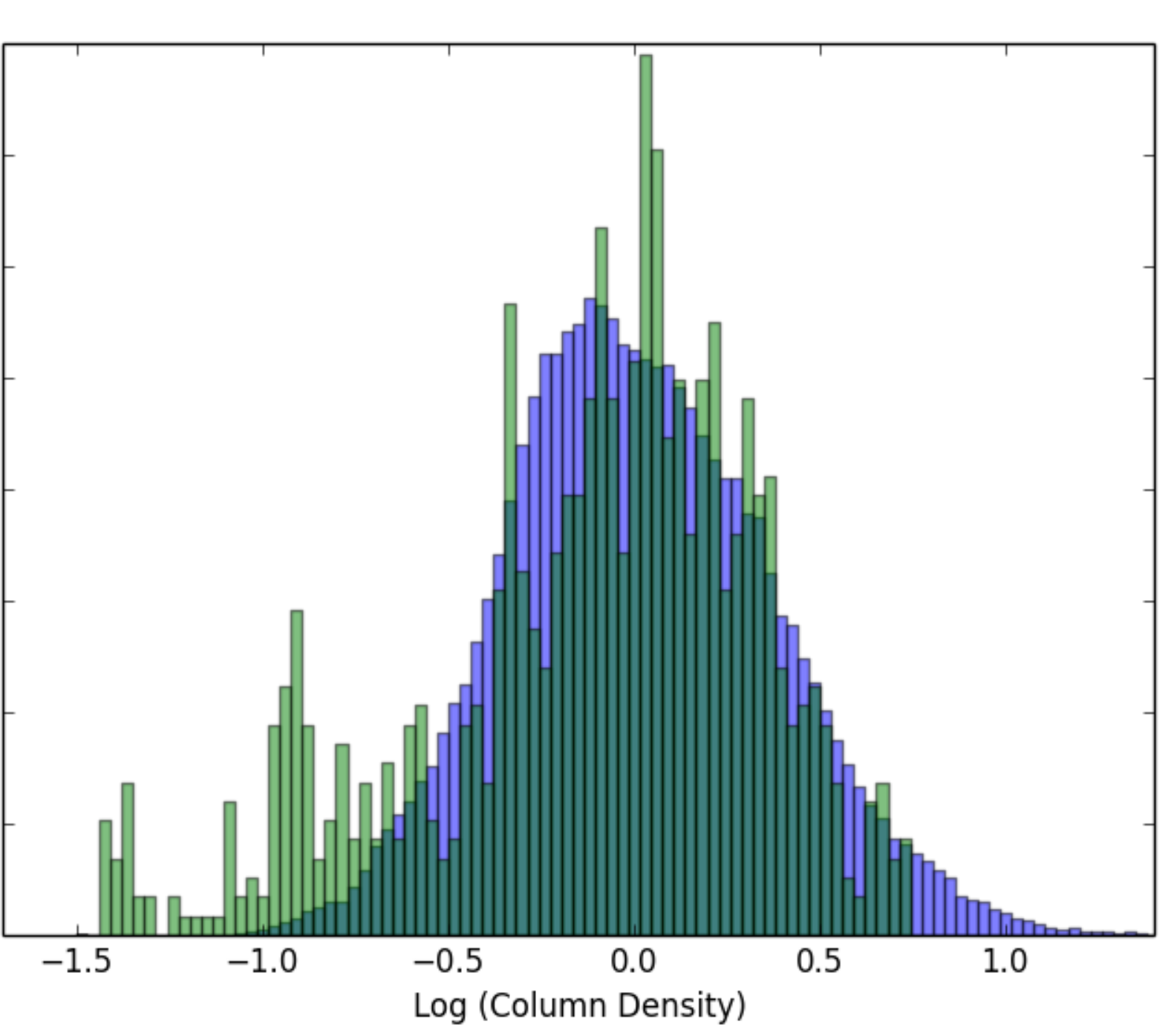}
\caption{Distribution of radial column densities at $t = 7 t_{tr}$, comparing the three-dimensional simulation (in blue) to the
corresponding two-dimensional simulation (in green).}
\label{fig:column3}
\end{center}
\end{figure}

\begin{figure*}[!hbtp]
\begin{center}
\includegraphics[width=6.in]{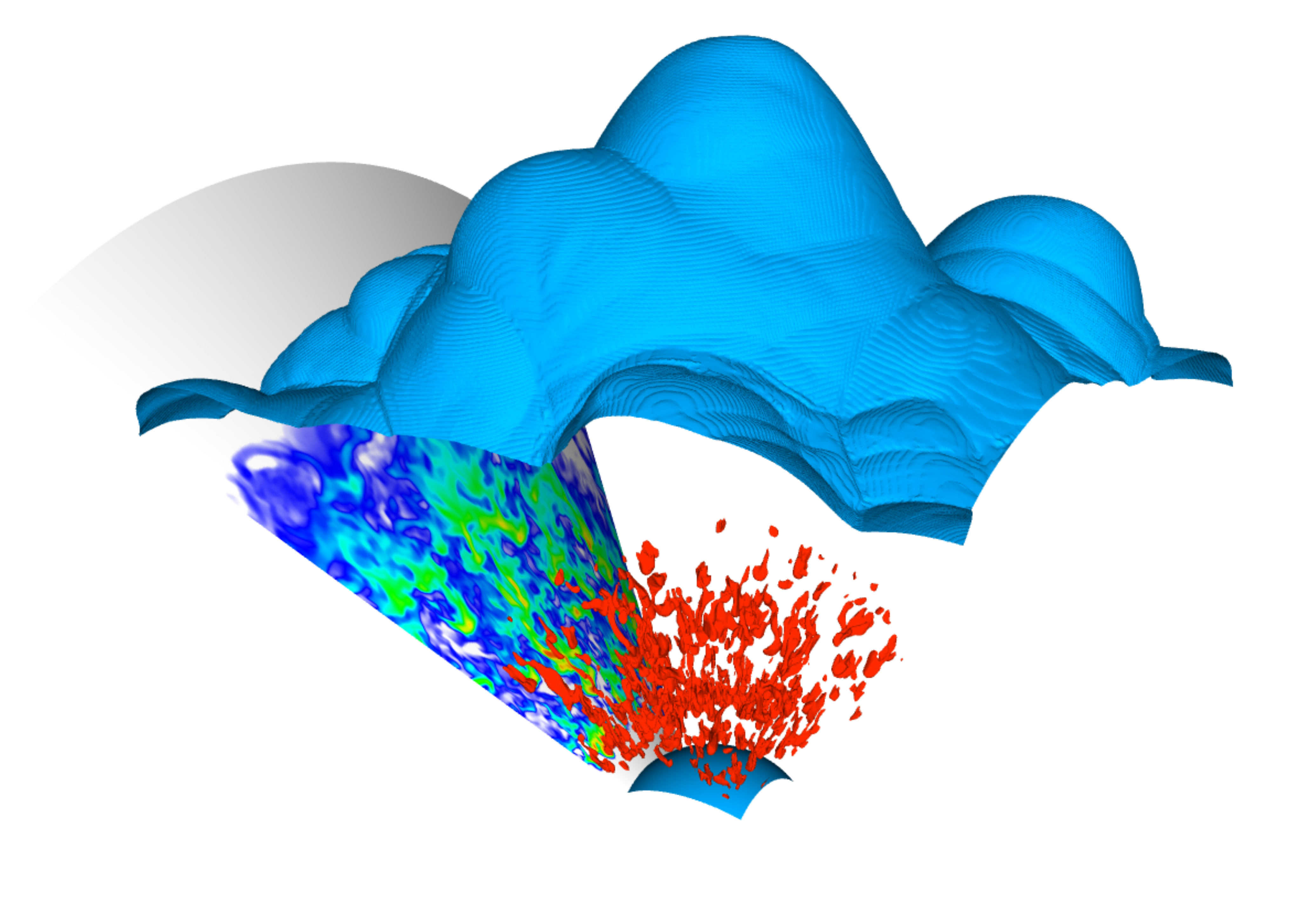}
\caption{The bulging structure of the forward shock in a three-dimensional simulation is illustrated here
with a surface of constant pressure, shown in blue.  One of the side walls of the simulation domain is colored by gas
density as in Figure 7.  The highest density clumps of shocked ejecta are shown as red surfaces.
}
\label{fig:shock3d}
\end{center}
\end{figure*}

\section{COMPARISON WITH PREVIOUS STUDIES OF BLOWOUT}

There have been 2 previous studies of pulsar nebula expansion into the blowout regime.
\cite{chen16} simulated the interaction with a realistic, exploded stripped star and continued the evolution to interaction with a steep ejecta density profile.
The simulation of the 1 ms magnetar by \cite{chen16} has higher resolution than our simulations, but they stopped their simulation
shortly after the break out phase was initiated. 
Although there are differences between their simulations and ours on small scales, the larger scale structure is similar.
\cite{chen16} find highly turbulent motions in the late phases of their simulation and conjecture that they are due to the
nonlinear thin shell instability (NTSI) \citep{vishniac94} operating between the pulsar wind termination shock and the outer shock front.
We do not find evidence for the NTSI in our simulations.
The termination shock is nonradiative and does not lead to a thin shell.  The absence of a spherical termination shock in the late time
simulations of \cite{chen16} may arise from the injection of the pulsar luminosity.  Our models inject this luminosity as kinetic energy
in a supersonic wind at small radii, while \cite{chen16} inject the pulsar luminosity as thermal energy at small radii.  At early times 
in their simulations the
expansion of gas from the injection region leads to a supersonic wind and the formation of a quasi-spherical termination shock as in
our simulations. At late times the RTI fingers push the termination shock to smaller radii and eventually past the transonic region.  At this
point the termination shock vanishes and the expansion of the pulsar wind is entirely subsonic with respect to the expanding shell and RTI structures.

The simulation of \cite{suzuki16} should be more directly comparable to our work because they adopted a two power law density structure
for the freely expanding ejecta as we did.
They have $\delta =1$ and $n=10$, which implies expansion with $\eta=1.25$ instead of $\eta=1.2$ in our $\delta=0$ model.
The larger acceleration could give somewhat stronger RTIs, but we expect qualitatively similar results.
The time for the shock wave to reach the density transition, $t_{tr}$ here, is $t_{br}$ in \cite{suzuki16}.
In \cite{suzuki16} there is an initial self-similar expansion law with $\eta =1.25$ as expected.  However,
the appearance of the density structure of the unstable gas is different from our results.
 \cite{suzuki16} find a higher degree of asymmetry in the forward shock front at a time just before $t_{tr}$ (see their Fig.\ 1).
 It is unlikely that the difference can be explained by the different value of $\eta$.
  \cite{suzuki16} start their simulation at $t=0.02 t_{tr}$ and so their simulation is unlikely to have evolved to the fully
  developed RTI by $t_{tr}$.
  They inject power in a region 10 zones in radius in a Cartesian grid, which could introduce asymmetry.
 Also,  \cite{suzuki16} do not use an expanding mesh, which limits the resolution in the early phases.
 
 The asymmetry in the early evolution of the  \cite{suzuki16} simulation carries over into the blowout phase.
 The early evolution leads to low density bubbles extending out near the symmetry axis; this may occur because of
 the high density finger along the axis of symmetry, noted in Section 2.
 When the flow gets to the blowout phase, the break through of the shell occurs more easily in the off-axis directions, leading
 to channels through ejecta gas in which the shocked pulsar wind gas flows.
 Our simulations also show rapid flow through channels, but there is no special orientation relative to the symmetry axis.
 Bulges in the forward shock wave resulting from the channel flow are present in our simulations and that of \cite{suzuki16}.
 \cite{suzuki16} used a special relativistic code and assumed the injected power had a baryon richness $\Gamma_{cr}=20$.
 The flow involving the shocked pulsar wind is generally non-relativistic until the break out of the shocked wind occurs and
 internal energy in the shocked wind is converted to kinetic energy.
 \cite{suzuki16} find that the channeled flows  reach Lorentz factor $\Gamma \sim 5$.
 We use a non-relativistic code, but allow for a high velocity wind and find that the outflows lead to a qualitatively similar structure
 to that found in the simulation of \cite{suzuki16}.

\section{DISCUSSION}

Our simulations show that there is a qualitative difference in the development of the RTI between the early and late phases of the
pulsar nebula evolution.
In the first phase, the swept up shell does form filaments, but the bubble is contained by the ram pressure of the external
supernova ejecta.
In the second phase, the ram pressure of the ejecta is not able to contain the bubble and it bursts through.
To reach the second phase, the energy deposited from the pulsar spindown must be  larger than the 
kinetic energy in the supernova ejecta.
Thus, this phase is relevant to cases where the supernova has an especially low energy or where the rotational energy
of the pulsar is especially large.

\subsection{Crab Nebula}

Although it has long been recognized that the energy in the observed Crab Nebula is small for a supernova, about $10^{50}$ ergs,
it is only recently that the evolution of a pulsar nebula in a low energy supernova has been examined as a model for
the Crab \citep{yang15}.
One argument for the low energy nature of the Crab ejecta is that much of the expected mass of the progenitor star is in
the ejecta swept up by the pulsar nebula.
In fact, much of the mass is in a set of dense filaments that are interior to the outer edge of the Crab.
There is evidence for an inner shell of filaments that is at about half of the radius of the outer edge \citep{clark83};
interior to the shell, there is primarily just synchrotron emission.
This structure does not agree with that expected during the first phase of evolution, when the denser filaments are
in the outer part of the  nebula and the density gradually decreases toward the central regions (Fig.\  \ref{fig:self-sim}).
On the other hand, during the second phase, the dense shell accumulated during the first phase gets left behind as
the low density bubble matter breaks through the supernova ejecta (Fig.\  \ref{fig:blowout}).
The initial shell of material forms an inner shell inside of which is little ejecta.
These properties are also shown in observations of the Crab Nebula, indicating that it has evolved toward the second phase.

\cite{yang15} discussed the possibility that the  northern ``jet'' in the Crab \citep[e.g.,][]{gull82} is related to the
blowout phenomenon.   The jet is a source of nonthermal radio emission, showing that the bubble gas has pushed out in that direction.
The simulations presented here do not show any features that resemble the jet, which might require allowance for
a magnetized flow.

\subsection{Energetic Magnetars}

In the magnetar model for superluminous supernovae,  the magnetar
bubble expansion may be out to the steep power law density decline \citep{kasen10,kasen15}.
\cite{chen16} and \cite{suzuki16} addressed the importance of the blowout phase for this model; we make some brief comments here.

The considerations of Section 2 can be used to estimate whether the supernova enter the blowout phase.
\cite{nicholl17} have recently modeled the multi-color light curves of SLSN-I in a systematic way.
They found that the 1-$\sigma$ range of initial spin periods is $1.2-4$ ms, which is a range relevant to the blowout phenomenon.
\cite{nicholl17} found the range of ejecta kinetic energies to be $1.9 - 9.8 \times 10^{51}$ ergs, but these high values refer to
the time after the ejecta have been accelerated by the magnetar power and not to the initial supernova energies.
The supernova energy is typically taken to be $1\times 10^{51}$ ergs, but there is not direct observational evidence for this value.

In the magnetar theory, the first observable effect of the pulsar nebula is the breakout of the shock front driven by the pulsar bubble.
\cite{kasen15} have discussed the breakout signature for the standard magnetar case with the intent of explaining the optical
precursor emission that is observed for some superluminous supernovae.
However, the luminosity of the shockwave in the stellar ejecta is only 1.5\% for the case of expansion in constant
density ejecta  \citep{chevalier92} and the
breakout emission does not distinguish itself from the main peak of the light curve unless there is incomplete thermalization
in the shocked wind bubble \citep{kasen15}.
The strongest signal is obtained when $t_{tr}< t_p$ and the breakout occurs when the outer shock front is in the
steep part of the density profile \citep{kasen15}.
These are the conditions most favorable for the RTI leading to blowout.
The escaping material from the blowout can drive a faster shock wave into the surroundings, thus producing a more prominent breakout signal.

The main light curve in the magnetar model is produced by the diffusion of radiation from the shocked pulsar wind bubble.
The effect of the blowout is to allow the more rapid expansion of the bubble contents and thus more rapid adiabatic expansion losses.
The loss of internal energy should lead to a decline in the light curve below what is expected in the absence of instabilities.
The blowout can also hasten the radiative losses of the gas.

\cite{metzger14} considered the breakout of an ionization front through the shell of ejecta.
If the material surrounding the pulsar bubble could be completely ionized, X-ray emission from the shocked bubble could escape and
be observed, as possibly occurred in the case of SLSN-1 SCP 06F6  \citep{levan13}.
X-ray emission could occur if there is  a pair plasma in the shocked pulsar wind.
The action of the RTI would  enhance the possibility of the escape of ionizing radiation.

\section{CONCLUSIONS}

When a pulsar or magnetar is born in a supernova, the bubble of relativistic particles and magnetic field created by the pulsar
expands in the freely expanding supernova ejecta.
The supernova ejecta are swept into a shell that is accelerated and thus subject to Rayleigh Taylor instabilities (RTI).
While the shell is in the inner flat part  or the ejecta, Rayleigh Taylor fingers of ejecta extend in toward the pulsar, but
most of the ejecta mass remains in the shell and outer part of the nebula.
The ram pressure of the preshock ejecta prevents bubble gas from moving out into the ejecta.
The overall structure is self-similar as the shell expands.
Because of the instability, the shock moves more rapidly in the 2-D case than the 1-D case, but only by a factor of 1.07.
The pulsar wind termination shock is not significantly perturbed by the RTI.

If the pulsar power persists past the time that the outer shock arrives at the transition to the steeply declining supernova
density profile, the ram pressure of the ejecta can no longer contain the ejecta and a vigorous RTI leads to the
blowout of the bubble gas.
Although there is a self-similar solution for this phase in 1-D, the multi-dimensional flow is not self-similar.
The bubble gas escapes through channels that are created in the supernova ejecta.
This situation may apply to the Crab Nebula, where there are massive filaments inside of the outer edge,
and to superluminous supernovae if they are powered by millisecond magnetars.

\acknowledgments
We are grateful to the referee for a helpful report and to C. Fransson, K. Maeda, and X. Tang for comments on the manuscript.
This research was supported in part by NASA grant  NNX12AF90G.

%=====================================================================
\end{document}